# Symmetric Domain Segmentation in WS$_2$ Flakes: Correlating spatially resolved photoluminescence, conductance with valley polarization.

*Arijit Kayal,\* Prahalad Kanti Barman, Prasad V. Sarma, M. M. Shaijumon, R. N. Kini and J. Mitra\**

School of Physics, IISER Thiruvananthapuram, Kerala 695551, India

KEYWORDS

WS$_2$, Domain segmentation, Defects, Optoelectronic heterogeneities, Valleytronics

ABSTRACT

The incidence of intra-flake heterogeneity of spectroscopic and electrical properties in chemical vapour deposited (CVD) WS$_2$ flakes is explored in a multi-physics investigation, via spatially resolved spectroscopic maps correlated with electrical, electronic and mechanical properties. The investigation demonstrates that the three-fold symmetric segregation of spectroscopic response (photoluminescence and Raman (spectral and intensity)), in topographically uniform WS$_2$ flakes are accompanied by commensurate segmentation of electronic properties e.g. local carrier density and the differences in the mechanics of tip-sample interactions, evidenced via scanning probe microscopy phase maps. Overall, the differences are understood to originate from point defects, namely sulphur vacancies within the flake along with a dominant role played by the substrate. While evolution of the multi-physics maps upon sulphur annealing elucidates the role played by S-vacancy, substrate-induced effects are investigated by contrasting data from $WS_2$ flake on Si and Au surfaces. Local charge depletion induced by the nature of the sample-substrate junction in case of $WS_2$ on Au is seen to invert the electrical response with comprehensible effects on their spectroscopic properties. Finally, the role of these optoelectronic properties in preserving valley polarization, affecting valleytronic applications, in $WS_2$ flakes is investigated via circular polarisation discriminated photoluminescence experiments. The study provides a thorough understanding of spatial heterogeneity in optoelectronic properties of $WS_2$ and other two dimensional transition metal chalcogenides, which are critical for device fabrication and potential applications.

INTRODUCTION

Two dimensional transition metal dichalcogenides (TMDCs) have lately been the hotspot of materials research. Along with a readily tunable bandgap[1-3] that renders them insulating, semiconducting or metallic, TMDCs provide a unique platform for strong light-matter interactions like valley polarization,[4, 5] second harmonic generation,[6] optical Kerr effect,[7] electroluminescence[8, 9] etc. seldom realized together in conventional materials. The layered structure of these 2D systems, held together by van-der-Waals forces evidences strong Coulomb interactions between electrons and holes, nucleating a variety of quasiparticles, excitons, trions etc.,[10-12] which have made the TMDCs attractive for investigating fundamental physics and exploring novel nanoelectronic and nanophotonic applications. However, the majority of the exotic results and reproducible device prototypes have been based on mechanically exfoliated TMDC flakes, which inherently lacks scalability. Chemical vapor deposition (CVD) grown TMDCs seem to offer a scalable alternative though reproducibility remains a challenge. In general, the optical, electrical and optoelectronic properties of CVD grown flakes show a fair degree of growth parameter-dependent variability. Not only between flakes but often within single flakes,[12-22] resulting from the complex interplay of differences in local disorder, stoichiometry, doping and strain, which are nucleated by the exact conditions of the growth process i.e. abundance of chemical species, nature of the substrate, temperature etc. However, beyond such generalized correlations, establishing an exact causal relationship and control between the growth parameters and the resulting physical properties are non-trivial. Therefore, establishing spatial correlations between the local variations in optical, electronic and spectroscopic properties of such systems is vital for their applications and benchmarking ranges of acceptability. Comprehending intra-flake heterogeneities assume added significance since the physical properties in the TMDCs are strongly influenced by their surfaces and interfaces, which sustain a large fraction of defects as dictated by the thermodynamics of the growth process. One of the most effective ways of evidencing intra-flake heterogeneities are spatial photoluminescence (PL) maps,[13, 15-17, 20, 22] often evidenced as symmetric, segmented domains of high- and low-intensity PL. It's worth mentioning, that the domain segmentation discussed here is distinct from reports of polymorphism and coexisting 1T (metallic) and 2H (semiconducting) phases observed in TMDC flakes.[7, 23, 24] Semiconducting TMDCs like $WS_2$, as investigated here show a non-centrosymmetric crystal structure with a hexagonal Brillouin zone and multi valleyed band structure. For monolayers the smallest energy gap between the conduction and valence bands occur at their K and K′ points, evidencing a direct bandgap, which changes to indirect bandgap for multilayer $WS_2$.[1, 25] The strong spin-orbit coupling due to the heavy metal ions[26] lifts the degeneracy at the K and K′ points, which coupled with the strong Coulomb interactions stabilizes different excitonic bound states for spin -up and -down electrons. Consequently, the PL spectrum is a convolution of multiple excitonic decay channels, along with emission from trions and defect state related de-excitation. Thus any spatial variation in PL would originate from local differences in the relative contribution of the above components. However, since the PL spectra are strongly influenced by both intrinsic parameters like the number of layers, local defects and strain, and extrinsic parameters like temperature, excitation wavelength etc., a wide range of responses have been reported, which are often difficult to reconcile and yield a generic picture of light-matter interactions in WS2

Typically, for CVD grown WS2, the flakes vary in size from few to tens of microns, with the segmented domain size varying between $10\ \mu m$ to $1\ \mu m$. The relatively large domain sizes enable investigation of domain heterogeneity through optical spectroscopy techniques, though their diffraction-limited spatial resolution restricts their scope in obtaining correlations between various physical properties, especially with local abundance of defects. For the domain segmented samples investigated here, the domains in the PL maps appear in triangular shape, arranged with 3-fold rotational symmetry around the flake and are discriminated by alternating high and low emission intensities, designated as $\alpha$ and $\beta$ domains respectively.[15, 17, 18, 22] Cumulatively, the previous reports indicate that the contrast in the PL maps primarily originates from the intrinsic tungsten vacancies (WV) and sulfur vacancies (SV) by correlating spatial variation of defect type and density, from high-resolution TEM with local PL intensity.[13, 15, 22] However, the results also establish strong correlation with local electron density and atomic termination at the edges of the flakes and there remains significant differences between the proposed explanations, some of which are not commensurate with each other, especially while consolidating the electrical, electronic and optoelectronic response of such systems. Spatially resolved Raman studies employing the first-order Raman modes (in-plane: $E_{2g}^1(\Gamma)$, out of plane: $A_{1g}(\Gamma)$) and the second order modes ($2LA(M)$ and $LA(M)$) have enunciated the role of lattice differences, defect landscape and strain in determining the differences between the domains. Again significant variations between the observations[15, 22] have hindered the emergence of a unified picture, further complicated by the observations[20] that evidence the role of the substrate and sample delamination in determining the physical properties of TMDCs.[27-29] Atomic force microscopy (AFM) based techniques have been used to probe the heterogeneities at higher spatial resolution primarily yielding topographic and phase-contrast information. The associated techniques also allow correlated measurements of local electrical, electronic and optoelectronic properties, which will further the present understanding of the origin and functional merits of the localized heterogeneities. Again, previous investigations using conducting AFM (CAFM) have yielded observations that are various and contradictory. While some have conflated higher defect density with conductivity and quenching of direct bandgap emission[17, 30] others have reported the opposite e.g. the highly emissive edge of triangular flakes carry a high local current.[31] Further, near field techniques like scanning near field optical microscopy (SNOM) and tip-enhanced Raman scattering (TERS) have provided insights into the role of defects in determining the energetics of the system.[32-35] Overall, these scanning probe results show that charge impurities,[29] surface roughness,[36, 37] proximity of the substrate and local strain[33, 37] play a critical role in the optoelectronic properties of these 2D materials that need to be explored further.

Comprehending the luminescence properties with respect to the band structure, the defect landscape, and associated parameters like carrier density and effect of the substrate assumes added significance while exploring valleytronics. Exploiting valleytronics requires the ability to selectively control specific valley populations aided by long valley polarization lifetimes ($\tau_v$) that are required to be substantially longer than any operational timescale of the device e.g. gate modulation time. As reported before[38-42] valley polarization lifetime is a convoluted parameter influenced by band structure, point defects and the carrier density of the system.

Against this backdrop, we have investigated spatial heterogeneity and domain segmentation in single WS$_2$ flakes with a range of coupled electrical, electronic and optical probes on as-grown samples on Si, as well as those transferred onto polycrystalline Au substrates. In this multi-physics investigation, the symmetric domain structure observed in WS$_2$ flakes is mapped by spatially resolved PL and Raman spectroscopy in conjunction with spatially resolved electrical, electronic and optoelectronic investigations, using a CAFM coupled with optical excitation. Correlation among these various maps along with topography and phase map elucidates the role of defects, strain, and substrate in deciding the optoelectronic properties of the flakes. Our results demonstrate that for samples on Si, the high luminescent domains are accompanied by higher local conductance, understood to originate from structural defects. The presence of the defects also results in the spectral redshift of the PL emission[11, 14, 43] and appearance of strong $LA(M)$ Raman mode[44] of WS$_2$, selectively within the high conducting domains. Difference between the samples' response, subsequently annealed in vacuum and in sulfur vapor provide evidence that quenching of SV leads to near-complete disappearance of the characteristic domains, evidenced across all the correlated multi-physics maps, which is contrary to some of the earlier reports.[19, 21, 30] Investigating the effects of physical transfer and the role of a highly conducting substrate (Au) on the local optoelectronic properties of the flakes show interesting differences from their as-grown counterparts, notably darker PL domains are now associated with relatively higher local conductivity. While some of these differences may be attributable to strain-induced during the transfer process, the exact role of the Au surface in modulating the luminescence properties still remains unclear. Further, helicity resolved PL (HRPL) experiment that yields the degree of circular polarization (DCP) parameter on WS$_2$/Au provides further insights into the relative propensity of direct radiative recombination vis-à-vis valley scattering, likely through modification of the exciton decay and intervalley scattering lifetimes. It is anticipated that this cohesive multi-physics investigation, mapping spectroscopic, electrical, electronic and optoelectronic properties on these segmented domains will not only contribute to a better understanding of the spatial heterogeneities in determining the optoelectronic properties of these CVD grown WS2 samples but also contribute to better design and exploitation of the same for future device applications.

RESULTS AND DISCUSSIONS:

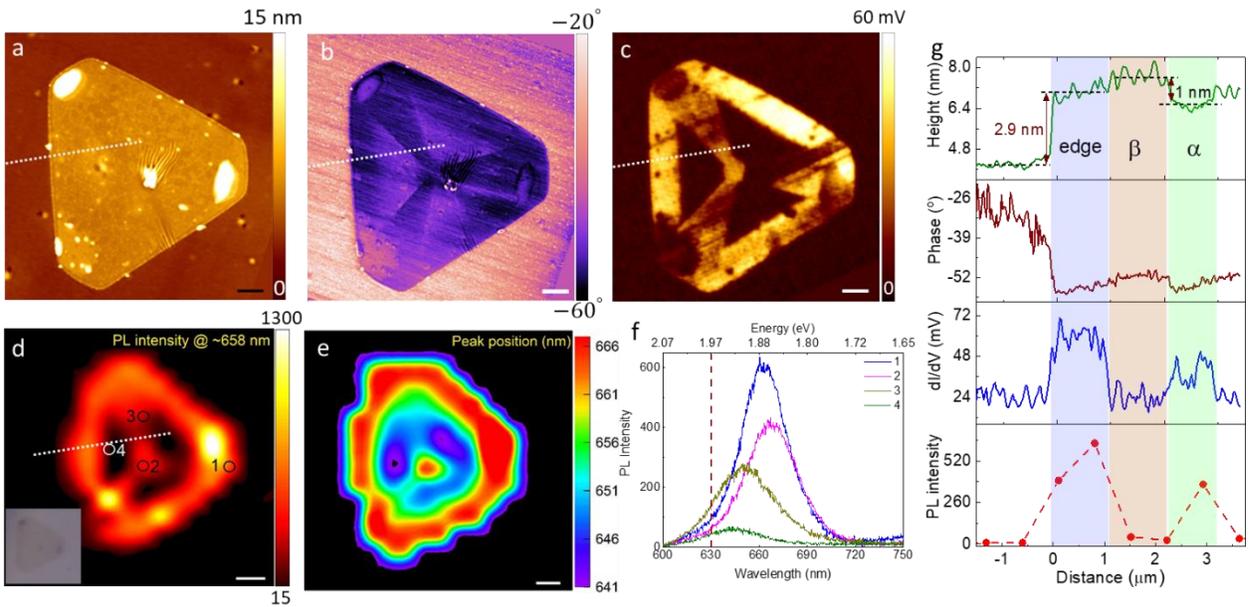

**Figure 1.** (a) Topography and (b) Phase contrast map of a $WS_2$ flake on Si (sample S1) acquired with tapping mode AFM. (c) Spatially resolved local conductance ($dI/dV$) map, recorded at $+10\ V$ simultaneously with topography using C-AFM. Spatially resolved map of PL (d) intensity (inset shows optical image of the flake), (e) peak position of the flake shown in figure a, recoded with 532 nm excitation. (f) PL spectra at the different locations marked in figure d. The dashed red line shows the energy of excitonic decay corresponding to a direct bandgap (g) Correlated line scan along the white dotted line in figures (a-d), showing the variation of relative height, phase, $dI/dV$ and PL intensity. All scale bars: 1 $\mu m$.

WS$_2$ flakes varying from one to a few monolayers in thickness were grown on Si substrates using CVD,[5, 45] (see methods section for details) and are discussed here. While the majority of the as-grown flakes presented homogeneous physical and spectroscopic properties (figure S1 in supplementary information), some show signatures of 3-fold symmetric domain segmentation. Figure 1a shows the AFM topography of an as-grown WS$_2$ flake (sample S1) with a nominal thickness of $2.9\ nm$, with a relatively uniform topography. The corresponding phase map, (figure 1(b)) shows phase contrast between alternating high and low phase domains with 3-fold symmetry, which is similar to that observed in the PL intensity map (figure 1d). Though the PL map discriminates the alternating high ($\alpha$) and low ($\beta$) intensity domains it lacks the spatial resolution of the phase map but depicts a continuous bright edge[12, 14, 43] that is not well differentiated in the phase map. Further, the PL intensity variation is accompanied by a symmetric variation of PL peak emission wavelength, as plotted in figure 1e. Across the domains the PL peak position varies from $650\ nm$ ($1.9\ eV$) at the low-intensity regions to $670\ nm$ ($1.85\ eV$) at the high-intensity regions, with linewidths in the range of $100 - 150\ meV$. Unlike the spatial intensity variation[12, 14-22] the corresponding $\alpha\beta$ domain segmented PL peak shift has not been reported or examined in detail. Figure 1f plots the PL spectra at various locations (numbered in figure 1d) which along with figure 1e shows that the spectral maxima from the center and the edges are brighter and more red-shifted compared to emission from the interior, especially from the darker $\beta$ domains. Spectra from the $\beta$ domains have maxima closest in energy to that corresponding to direct band gap excitonic decay in WS$_2$.[1, 3, 25] These observations broadly follow earlier reports and reiterate the important role played by defects and manifest in the characteristic spatial variation of the PL properties. These characteristics are either mediated via the defect states[11, 12] or by inducing local carrier doping that favor trion formation,[3, 18] or locally straining the lattice etc. or a combination thereof.[46] Variation in the local carrier concentration would also be reflected in the local electronic and electrical properties, consequently Figure 1c shows the spatially resolved CAFM $dI/dV$ (conductance) map for a sample bias $(V_b)$ of $+10\ V$. The contrast represents the variation in the local conductance scaled to an output voltage in the range 0 to $60\ mV$. The conductance map again exhibits the 3-fold symmetric $\alpha$ domains observed in the phase and PL intensity maps, along with a peripheral high conductance band, interleaved with the three $\beta$ domains of lower local conductance. Note, the irregular low conductance patches at the truncated corners, which likely originate from debris, evidenced as bright blobs in the topography (figure 1a). Similarly, the presence of wrinkles at the center and middle of the lower edge of the topography are reflected as low conductance regions in the conductance map. The coincidence of the high PL intensity with red-shifted emission and high conductance gives further credence to their common origin i.e. nature and abundance of local defects.

The conductance and phase maps better resolve the cross-over between the $\alpha$ and $\beta$ domains, compared to those acquired by optical techniques. The phase maps bear testament to the local tip-sample adhesive properties and quantify energy dissipation of the AFM cantilever due to tip-sample interaction. They carry information regarding the chemical heterogeneity across the surface,[19, 47, 48] thus, the high relative phase shift in the $\alpha$ domains compared to the $\beta$ domains, indicates stronger tip-sample adhesion that stems from the systematic variation of surface defects across the domains. The presence of defects at the sample surface can give rise to unsaturated dangling bonds leading to relatively stronger adhesion than defect-free domains. The highly correlated nature of all the above probes, which otherwise originate from distinctly different physical interactions on the sample is further elucidated in the correlated line scans (figure 1g), showing height, phase, conductance and PL intensity variation along the white dashed lines in figures 1(a-d). The three color-coded regions, blue, orange and green correspond to the edge, $\beta$ region and $\alpha$ region respectively. Figure S2 in supplementary information shows the spatial maps of the various measurements discussed above on three flakes (samples S2, S3, S4), that exhibit domain segmentation, demonstrating the overall reproducibility of the individual observations and the correlated nature of the same. Across all the samples, in spite of the uniform sample topography, the PL, phase and conductance maps evidence 3-fold symmetric, triangular domains extending from the center to the truncated corners, which have higher conductance and PL intensity along with higher tip-sample adhesion, all connected to the type and abundance of local defects.

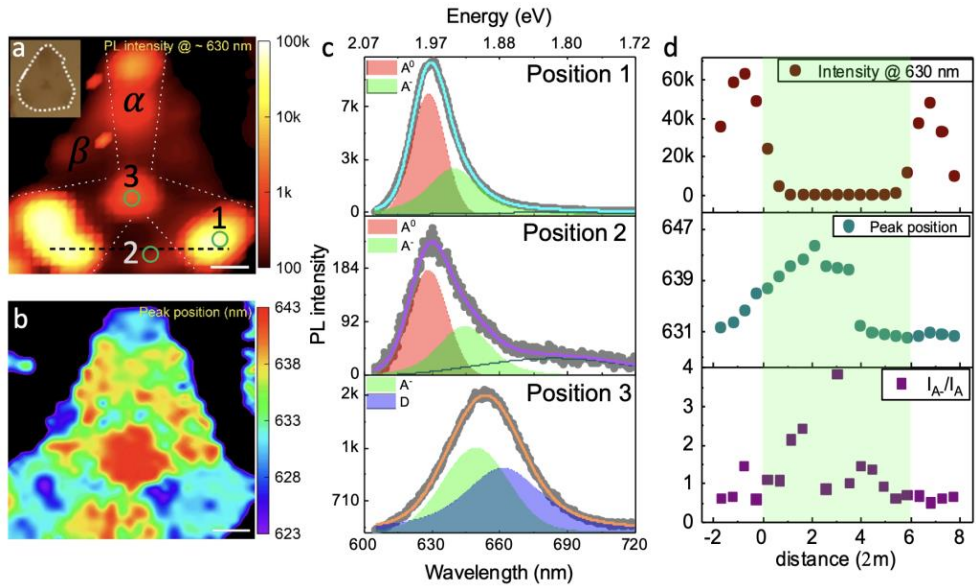

**Figure 2.** Spatially varying photoluminescence of a WS$_2$ flake on Au (sample S5) (a) PL map showing segmented domains (b) map of PL peak wavelength (c) PL spectra at points 1, 2 and 3 shown in (a), and (d) correlated plots showing the variation of PL intensity, peak wavelength, the intensity ratio of trion and exciton intensity, along the black dashed line shown in (a) Scale bar: 2 $\mu m$

Some of the as-grown 2D WS$_2$ flakes that exhibit domain segmentation were subsequently transferred to 50 nm-thick Au (with an under layer of 5 nm Cr) coated Si wafers using a wet-transfer method.[49] Figure 2a shows the spatially resolved PL map at 630 nm for one such WS$_2$ a flake (sample S5), with the inset showing the optical image. Akin to the PL intensity maps of the flakes on Si substrate, figure 2(a) shows domain segmentation, with the $\alpha$ domains (demarcated by white dashed lines) connecting the center to the truncated edges that have the highest luminescence. Figure 2b shows the corresponding map of peak PL wavelengths with variation in the range $620 - 640\ nm$ ($2.0 - 1.93\ eV$). Representative spectra from three points in figure 2a in the $\alpha$ and $\beta$ domains, and from the center are shown in figure 2c. Intensity of emission from the $\alpha$ domains are not only significantly higher than that from the $\beta$ domains, but are much higher than that recorded in the $\alpha$ domains of flakes on Si (figure 1). Further, emission from both the $\alpha$ and $\beta$ domains show maximum intensity around $\sim 630\ nm$ ($1.97\ eV$) in contrast to that for the flakes on Si, except for the center that evidences red-shifted emission spectrum maximizing around 660 nm. The elongated tail at lower energies in all spectra indicate presence of multiple emission channels, attributable to decay of neutral excitons ($A^o$), negative trions ($A^-$) and defect ($D$) state mediated emission. Spectral deconvolution quantifies the relative contribution of the channels to the total intensity as depicted in figure 2c. The best fit linewidths $60\ meV$ and $80\ meV$ of the exciton and trion fractions and their mean separation of $\sim 33\ meV$ matches well with previous reports,[10, 11] justifying their identification. Importantly, variation in PL intensity associated with the excitons and trions across the $\alpha$ and $\beta$ domains changes systematically as seen in figure 2d, which co-plots PL intensity variation along with peak emission wavelength and the $I_{A^-}/I_A$ ratio along the black dashed line shown in figure 2a, extending across $\alpha - \beta - \alpha$ domains. The weaker $I_{A^-}/I_A$ values across the $\alpha$ domains indicate lower relative contribution of trion decay to the overall PL intensity compared to the $\beta$ domains, which evidence relatively higher contribution from trions compared to excitons. This observation appears counterintuitive to the previous evidence identifying the $\alpha$ domains as negatively doped electron rich regions as reported before. Again, local increase in trionic to excitonic contribution has been attributed to excess carrier concentration induced by the defects[32, 50] and to substrate induced doping.[3, 27]

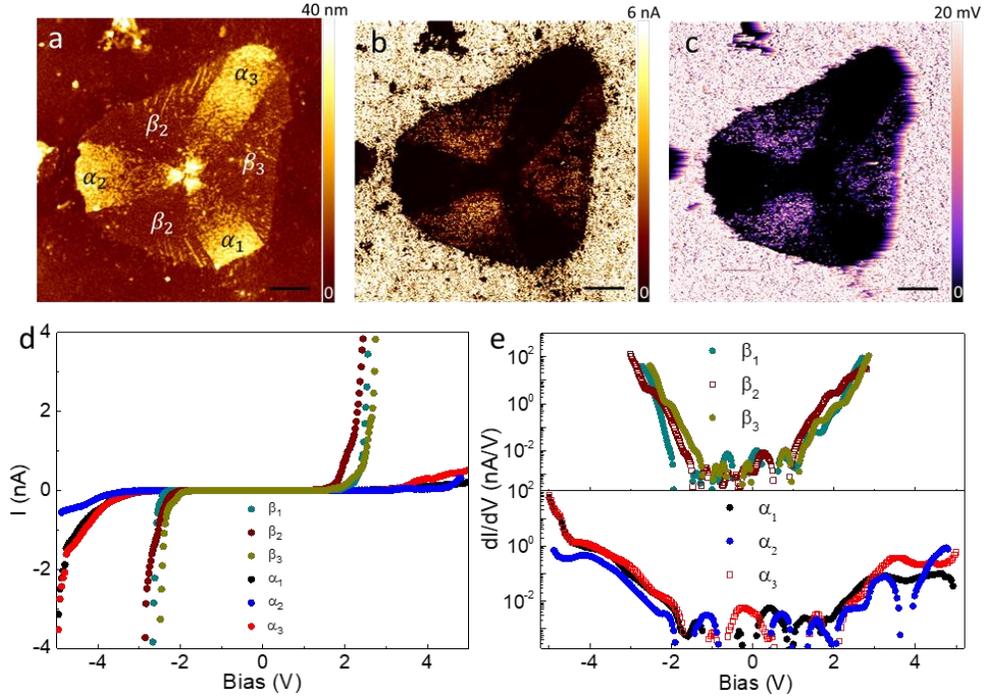

**Figure 3.** Spatial maps on WS$_2$ flake on Au (sample S5) (a) AFM topography, (b) CAFM current map (c) conductance map at +1.2 V and (d) point *IV* characteristics and corresponding (e) $dI/dV$ spectra at different locations in the $\alpha_{123}$ and $\beta_{123}$ domains identified in (a) Scale bar: 2 $\mu m$

Figure 3a shows the AFM topography of the sample S5 on Au, which now depicts 3-fold symmetric variation in height. The regions of $\alpha$ domains extending from the centre to the truncated corners have a significantly higher topography (~ 20 nm at the highest point) than the intervening $\beta$ domains, which are ~ 4 nm above the substrate as evidenced for flakes on Si. Appearance of height variation, within a single flake upon transfer though reported before[20] and explained as edge delamination effects [20, 51] remains surprising, due to the height difference between the $\alpha$ and $\beta$ domains ($\Delta h$ ~ 16 nm). The higher topography of the $\alpha$ domains originating from local growth of additional WS$_2$ layers is contraindicated by the high PL intensity since multilayer formation reduces PL intensity due to the indirect nature of bandgap of multilayer systems.[1, 14] The high correlation between the radial variation in PL intensity (figure 2a) and topography (figure 3a) within each $\alpha$ domain is also notable. In both cases they are high at the centre, decreases outwards and are high again at the edges. Unlike the topography of the flakes on Si, here it is also modulated with occasional nanobubbles and wrinkles along the edges. Further, the PL intensity contrast almost disappear between the regions of $\alpha$ and $\beta$ domains that have the same topographic height. Equally surprising are the CAFM measurements of current and conductance maps recorded at sample bias of $+1.2\ V$, which again evidence the 3-fold symmetry of segmented domains as shown in figures 3b and 3c. However, in both cases the $\alpha$ domains show lower current and local conductance compared to those in the $\beta$ domains, in contrast to that observed in the as-grown flakes on Si. The contrast in the local electrical properties is better quantified in the variation in the point *IV* characteristics (figure 3d) and the corresponding *dI/dV* spectra (figure 3e) obtained at various points across the segmented domains. Figure 4a shows a schematic of the CAFM measurement configuration i.e. a $WS_2$ flake lying on a Au substrate with the CAFM tip atop forming two junctions on the bottom and top surfaces of the flake. If the sample is homogeneous and the sample-substrate junction allows linear transport, the *IV* characteristics are typically dominated by electrical transport across the junction formed between the Au coated tip and the $WS_2$ sample. Perhaps both these assumptions are questionable in the present scenario. The single $WS_2$ flake is not homogeneous as far as electrical and electronic properties are concerned and if the height variation originates from edge delamination or even otherwise, the sample – substrate junction is also non-uniform and contributes to the observed current and conductance maps. In fact, the *IV* characteristics shown in figure 3d are representative of two back-to-back metal-semiconductor-metal Schottky junctions, which are asymmetric with applied bias. Assuming that the $WS_2$ flake is nominally *n*-doped, for -ve substrate bias the sample – substrate junction is reverse biased whereas the tip – sample junction is forward biased and vice versa for +ve substrate bias. Consequently, the overall *IV* characteristics will be dominated by the reverse bias characteristics of the top and bottom junctions under +ve and -ve bias, respectively. Figure S3 in the supplementary information shows a series of current and conductance maps acquired on the same flake at $-1.0\ V$, $+1.0\ V$, $+1.4\ V$ and $+2.0\ V$. At $-1\ V$ the current and conductance maps show little contrast between the domains, which become increasingly prominent under increasing positive bias. Even at $+2\ V$ bias, the outer edges of the $\alpha$ domains show practically zero current with conductance below the detection threshold of the current amplifier.

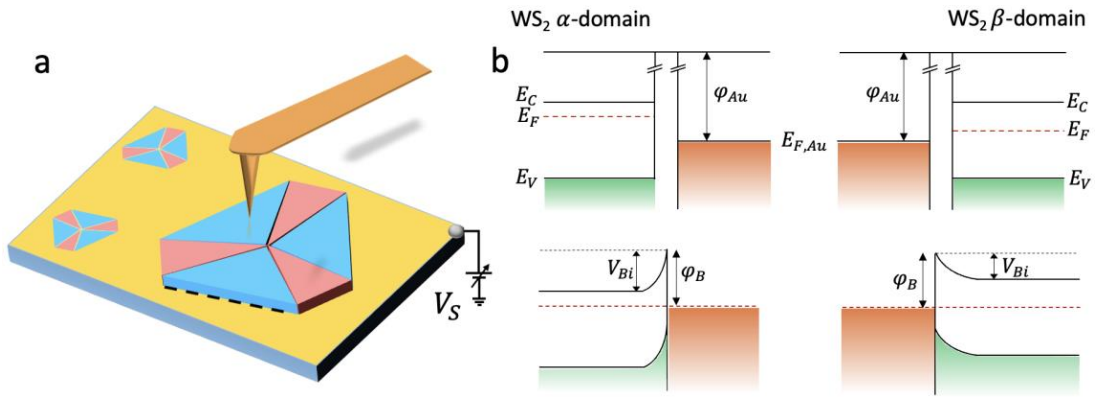

**Figure 4.** (a) Schematic of the conducting AFM measurement configuration showing the segmented domains in a WS$_2$ flake, on a Au substrate and Au coated CAFM tip on top. The dashed line denotes the substrate-sample junction (b) Energy band diagram of the junction between the $\alpha$ and $\beta$ domains of WS$_2$ with the Au substrate before and after junction formation.

In the $\alpha$ domain, the coincidence of intense PL, dominated by excitonic contribution and lower conductance points towards depletion in local electron density compared to that in the $\beta$ domains. To elucidate the statement further we refer to the band diagrams shown in figure 4b. The first and the second columns show the band alignment during contact formation between the WS$_2$ flake and the Au substrate in both the $\alpha$ and $\beta$ domain interfaces. In drawing the band diagram, it's worth noting that previous reports of WS$_2$ typically place the conduction band minima and the valence band maxima at $-3.9\ eV$ and $-5.8\ eV$ respectively[52] with the Fermi level ($E_F$) of the Au substrate around $-5.2\ eV$ and the later has free electron density far higher than that in either domain of WS$_2$. From the spatially resolved optical and electrical data for the WS$_2$ flakes on Si it may be extrapolated that in a free-standing WS$_2$ flake, with segmented domains, the $\alpha$ domains would have higher electron density compared to the $\beta$ domains, induced by local defects. Consequently, the former will have a higher $E_F$ than the latter i.e. $E_F(\alpha) > E_F(\beta)$, as shown in the first row of figure 4b. This is commensurate with Kelvin probe measurements on domain segmented WS$_2$ reported before, where the contact potential difference recorded on the $\alpha$ domains are higher than that on the $\beta$ domains.[17, 53] Thus, even within a free-standing WS$_2$ flake, there would be band bending in the azimuthal direction to equilibrate the differences in the local chemical potential between the 3-fold symmetric domains due to differences in local free electron density. However, once placed in contact with the Au substrate, that equilibrium would be replaced by a new one dominated by the chemical potential of the Au substrate. Since the absolute difference between the Fermi level of Au ($E_F(Au)$) and $E_F(\alpha)$ will be higher than that of the $\beta$ domain i.e. $|E_F(Au) - E_F(\alpha)| > |E_F(Au) - E_F(\beta)|$ band bending will be higher at the $\alpha$ domain – Au interface compared to the $\beta$ domain – Au interface. In effect the electron depletion in the $\alpha$ domain will be more compared to depletion in the $\beta$ domain. Finally, the new equilibrium will ensure that the Fermi level of both the $\alpha$ and the $\beta$ domains are pinned to that of the substrate as shown in the bottom panels of figure 4b. In the CAFM device configuration, where the Au coated AFM tip forms another Schottky junction on the top WS$_2$ surface, the more heavily electron depleted $\alpha$ domain evidences lower conductance compared to the $\beta$ domain. Thus, the regions of high and low current and conductance map contrast of the segmented domains for WS$_2$ flakes on Si are inverted for the flakes on Au. This also explains the progressively increasing contrast in the current and conduction maps with an increase in +ve bias, as seen in figure S3. Indeed, these observations have been replicated across multiple WS$_2$ flakes (samples S6, S7) transferred onto Au substrate, which displays domain segmentation, as shown in figure S4 of supplementary material and thus are understood to be generic in nature. This differential electron depletion across the $\alpha$ and $\beta$ domains also explain the variation in the PL spectra observed across the WS$_2$ flake. The $\alpha$ domains depleted of free electrons inhibits negative trion formation resulting in the PL spectrum being dominated by excitonic decay, relative to the $\beta$ domain.

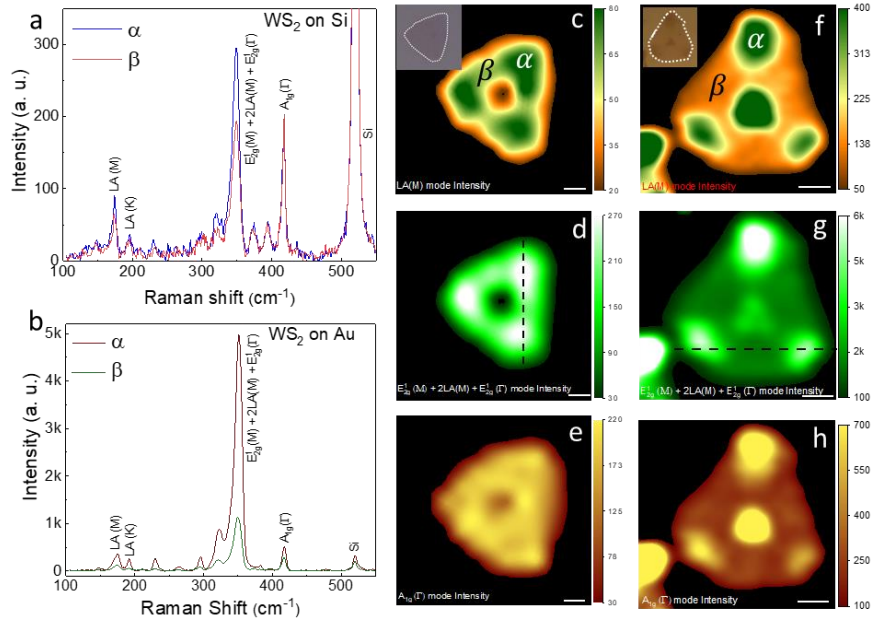

**Figure 5.** Spatially resolved Raman investigation. Raman spectra over $\alpha$ and $\beta$-domain on (a) $WS_2$ on Si (sample S4) and (b) $WS_2$ on Au (sample S5). Raman map of (c) $LA(M)$ mode, (d) $[E_{2g}^1(M) + 2LA(M) + E_{2g}^1(\Gamma)]$ mode, (e) $A_{1g}(\Gamma)$ mode for sample S4 (scale bar: 1 $\mu m$) and (f) $LA(M)$ mode, (g) $[E_{2g}^1(M) + 2LA(M) + E_{2g}^1(\Gamma)]$ mode, (h) $A_{1g}(\Gamma)$ mode for the sample S5. Inset of figure (c) and (f) shows the optical image of the flakes (scale bar: 2 $\mu m$).

Raman measurement has been used effectively not only to understand the lattice dynamics of 2D TMDCs but also to investigate the energetics of defects they harbor, the measurement being responsive to perturbations like adjacent layers,[54, 55] strain,[56, 57] and defects[44, 50, 58]. Figure 5a shows the Raman spectra acquired over $\alpha$ and $\beta$ domain of a WS$_2$ flake on Si substrate (sample S4) with 532 nm excitation. Both the spectra resolve the two characteristic peaks of the semiconducting 2H phase of $WS_2$ at $\sim 350\ cm^{-1}$ and $\sim 417\ cm^{-1}$. While the peak at $417\ cm^{-1}$ corresponds to the $A_{1g}(\Gamma)$ mode, which originates from the out of plane vibration of S-atoms, the peak at $\sim 350\ cm^{-1}$ is a convolution of three vibrational modes $E_{2g}^1(M)$, $2LA(M)$ and $E_{2g}^1(\Gamma)$, with the $E_{2g}^1(\Gamma)$ mode originating from in-plane relative motion of W and S atoms.[54] Both the $E_{2g}^1(\Gamma)$ and $A_{1g}(\Gamma)$ modes corresponds to scattering close to Brillouin zone center in contrast to the $2LA\ (M)$ mode which is a zone edge mode and arises due to 2-phonon scattering associated with resonant Raman excitation.[54] Two other modes, the $LA(M)$ and the $LA(K)$ modes also arise from second-order scattering processes, with the former being sensitive to lattice disorder and thus gives a measure of defect density, via an inverse square inter-defect distance dependence of its intensity.[44, 50, 58] Importantly, the acoustic phonons giving rise to the $LA(K)$ mode, resolved at $\sim 192\ cm^{-1}$,[59, 60] enhances intervalley $(KK')$ scattering responsible for valley depolarization.[61, 62] Figure 5b shows the Raman spectra on the $\alpha$ and $\beta$ domains of sample S5 (WS$_2$ on Au), whose PL and CAFM data are presented in figures 2 and 3. While the overall spectral composition of all the four spectra presented are similar, their differences accentuate while comparing the spatially resolved modal maps. The spatial maps of the $LA(M)$, $[E_{2g}^1(M) + 2LA(M) + E_{2g}^1(\Gamma)]$, and the $A_{1g}(\Gamma)$ modes are shown in figures 5(c-e) and 5(f-h) for WS$_2$ on Si and Au substrates. On both substrates, the $LA(M)$ and $[E_{2g}^1(M) + 2LA(M) + E_{2g}^1(\Gamma)]$ maps show the 3-fold symmetric variation of intensity, though each mode is more intense on the Au substrate. The localization of the high-intensity regions in the defect-related $LA(M)$ map with the $\alpha$ domains, in both the systems, further entrench the positive correlation of incidence of defects, enhanced PL and local conductance (see SI figure S5). The correlation between the maps of the strongest Raman mode ($[E_{2g}^1(M) + 2LA(M) + E_{2g}^1(\Gamma)]$) and PL is expected since higher intensity PL is indicative of stronger light-matter interaction that would benefit resonant Raman scattering processes like the $2LA(M)$ mode. The correlation is also evident in the synchronous line scans taken across both flakes, along the dashed lines extending across the $\alpha - \beta - \alpha$ domains (figures 5d & 5g) and shown in SI figure S5. Variation in intensity of the three dominant Raman lines are presented along with variation in height, conductance and PL intensity. Interestingly, only the $A_{1g}(\Gamma)$ maps (and line scans) discriminate between the flakes on the two substrates. While the corresponding modal intensity is almost uniform over the entire flake on Si (figure 5e and S5a), the map for the flake on Au (figure 5h) shows the 3-fold symmetry. In the latter, the intensity of this Raman mode corresponding to the out-of-plane vibrations is most prominent at the truncated edges of its $\alpha$ domains (figure S5b), which are also the regions with higher topography (figure 3a). This may be interpreted as weak interaction between the region and the underlying substrate and possibly delamination as discussed earlier. Local delamination of the $\alpha$ domain edges would also be commensurate with the lower conductance of the same regions evidenced in the conductance maps shown in figures 3(d), S3 and S5. However, the strain incurred via such a drastic height variation would question the physical integrity of the flake. Overall, the above results establish a strong spatial correlation between the local conductance, PL intensity and the Raman modal maps, all of which carry signatures of higher defect density in the $\alpha$ domains compared to the $\beta$ domains, within single WS$_2$ flakes. As mentioned earlier, previous reports have associated sulphur vacancies at the edge of WS$_2$ flakes with the incidence of higher intensity PL, especially with increasing contribution from trions due to higher local carrier densities and defect state mediated de-excitation. In both cases, the PL spectra are red-shifted compared to that of unperturbed samples suggesting higher defect density preferentially at the flake edge.[12, 14] To test the hypothesis that the domain segmentation may arise from sulphur vacancies, selected WS$_2$ flakes on Si were annealed at 650°C for 30 mins in a Sulphur-rich environment with Ar as the carrier gas.

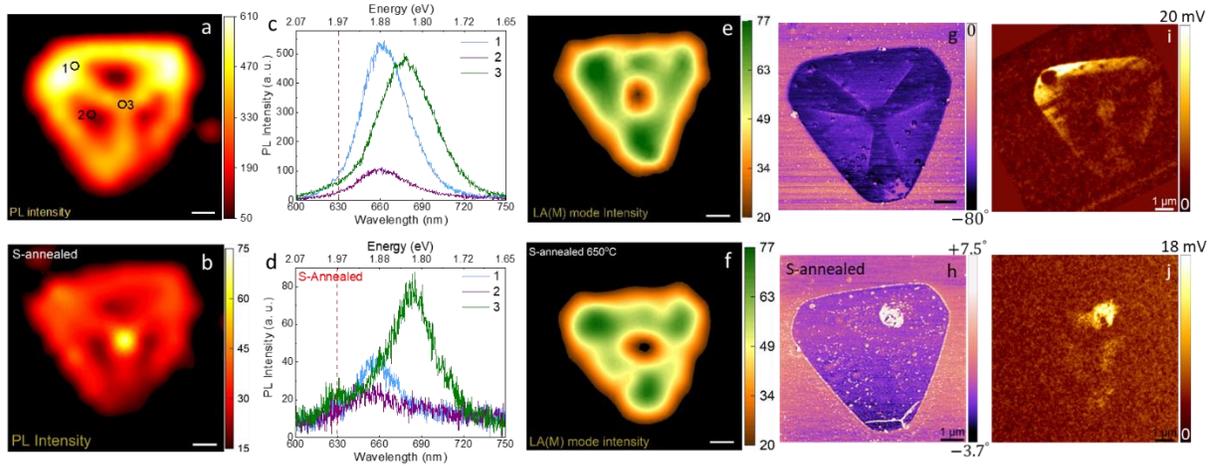

**Figure 6.** Spatially resolved maps and PL spectra of a WS$_2$ flake on Si (sample S2) (a-b) PL intensity (c-d) PL spectra (e-f) Raman $LA(M)$ mode (g-h) phase (i-j) conductance maps. Top row corresponds to as-gown sample and bottom row corresponds to post sulphur annealing.

Figures 6a and 6b show the PL intensity map of a WS$_2$ flake (sample S2, S2A) before and after annealing in a sulphur environment with the corresponding PL spectra recorded at three points (figure 6a) at the centre, and the $\alpha$ and $\beta$ domains shown in figures 6(c) and 6(d). Post annealing, the PL intensity contrast between the domains decreases significantly along with an overall reduction in the emission intensity across all regions. These observations are replicated across the other domain segmented WS$_2$ flakes (samples S1A, S3A, S4A) that were annealed, as shown in Supplementary figure S5. The spectra in Figure 6d also shows the appearance of a shoulder at around $630\ nm$ ($1.97\ eV$) that was absent prior to annealing (figure 6c), which is attributable to excitonic recombination. The drastic reduction in the intensity of the primary peak at $660\ nm$ after sulphur annealing indicates that the emission in the as-grown sample is indeed associated with S-vacancy in the system. Further, a comparison between the Raman maps of the $LA(M)$ modes (figures 6e and 6f), which are responsive to lattice disorder, also show weakening within the $\alpha$ domains, providing further evidence for the decrease in defects within the region. Perhaps the most distinct change in the flakes, brought about by Sulphur annealing is evidenced in the AFM phase maps. The 3-fold symmetry of the phase map prior to annealing (figure 6g) is lost in the annealed sample (figure 6h). Similarly, the conductance map also loses contrast upon annealing, and the flake becomes almost indistinguishable from substrate conductance, providing indirect proof that the original spatial contrast arose due to SV induced local carrier density variation. Overall, S-annealing elucidates the role of defects in realizing the spectroscopic, optical and electronic properties of these segmented films.

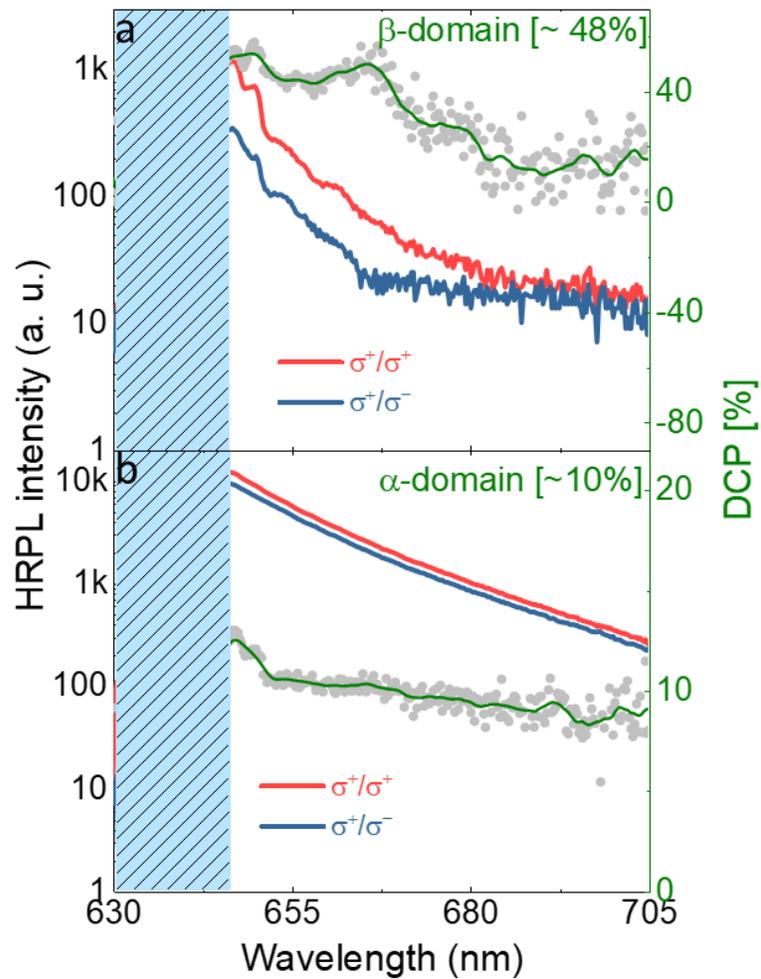

**Figure 7.** HRPL spectra obtained by using right circularly ($\sigma^+$) polarized, near-resonant (633 nm) excitation from the heterogeneous WS$_2$ flake (sample S5), and the corresponding DCP at RT for (a) $\beta$ and (b) $\alpha$ domains. The shaded region represents the stopband of the notch filter used to block the laser light from reaching the detector.

In order to explore the performance of domain segmented flakes for application in valleytronics, helicity resolved PL measurements were performed with CW laser excitation at $633\ nm$ (~$1.96\ eV$) and $532\ nm$ (~$2.33\ eV$), at RT. At both wavelengths, the samples were excited with right circularly polarized (RCP/$\sigma^+$) light and the polarization states of the emitted PL was discriminated as RCP or left circularly polarized (LCP/$\sigma^-$). The degree of circular polarization (DCP) parameter, as given below then quantifies the degree of valley polarization in the system,

$$DCP[\%] = \frac{I_{\sigma^+} - I_{\sigma^-}}{I_{\sigma^+} + I_{\sigma^-}} \times 100 \qquad (1)$$

Here, $I_{\sigma^+}$ and $I_{\sigma^-}$ are the RCP and LCP emission intensities. For the WS$_2$ flake on Au (sample S5) the overall PL spectrum is dominated by emission from the $\alpha$ domains with a peak near $630\ nm$. Figure 7 (a) - (b) shows the intensities of the RCP ($I_{\sigma^+}$) and LCP ($I_{\sigma^-}$) polarized PL along with the calculated DCP for $633\ nm$ excitation from the $\alpha$ and $\beta$ domains, respectively. The wavelength range $630 - 645\ nm$ (shaded region) has been removed since it represents the stopband of the notch filter used to block the laser excitation reaching the detector. The $\beta$ domains show DCP $\cong$ 50% while $\alpha$ domains show DCP $\cong$ 10% for the same RCP ($\sigma^+$) excitation, which is shown LCP ($\sigma^-$) excitation. The PL spectra and DCP for two other WS$_2$ flakes on Au (samples S6 and S7) are shown in Supplementary Information Figure S8, which are commensurate with the data presented above. The coincidence of higher DCP with low PL intensity in the $\beta$ domains and low DCP with high PL intensity in the $\alpha$ domains has been reported before.[40, 63] For excitation with RCP ($\sigma^+$) 532 nm light the PL and DCP obtained on samples S5 and S6 are shown in figure S9. Expectedly, the PL shows a single peak around 630 $nm$ but with negligible DCP for both the $\alpha$ and $\beta$ domains. Thus, excitation energy matching the luminescence peak strongly favors DCP in comparison to excitations at higher energies. Weak DCP alludes to high intervalley scattering such that electrons optically excited to the spin-up states, under $\sigma^+$ excitation, will be scattered to the spin-down states and decay giving $\sigma^-$ PL, and vice-versa. Thus the observed DCP also quantifies the relative strength of the two associated processes i.e. scattering rates of carrier recombination (e.g. excitonic decay rate) and the intervalley scattering and the DCP may be expressed in terms of the corresponding lifetimes as,[38]

$$DCP[\%] = \frac{P_0}{1+2\frac{\tau_e}{\tau_v}} \qquad (2)$$

Where, $P_0$ is the initial polarization, $\tau_v$ denotes the valley lifetime and $\tau_e$ denotes the exciton lifetime, including both radiative and non-radiative recombination processes. Equation (2) suggests that the DCP is increased by factors that decrease the $\tau_e/\tau_v$ ratio, either one or both of $\tau_v$ and $\tau_e$. Increase in $\tau_v$ due to higher carrier densities or short $\tau_e$ due to defects increase DCP, thus factors which influence the luminescence properties of TMDCs play a major role in deciding the DCP.[40-42, 63-65] Extrinsic parameters like the substrate and the excitation wavelength ($\lambda_{exc}$) have also been reported to significantly influence the observed DCP, as seen above in the difference between the DCP obtained with 532 nm (2.3 eV) and 633 nm (1.96 eV). Significant overlap between the excitation and emission line-widths increases DCP by disadvantaging intervalley scattering. While the difference in the overall DCP, in both the $\alpha$ and $\beta$, for the two excitations may be explained due to the difference in selectivity rules affecting the system, the difference in the spatial dependence of PL and DCP under the two excitation are non-trivial to reconcile. Firstly, for the same RCP excitation the ratio of $I_{\sigma^+}$ between the domains i.e. $I_{\sigma^+}(\alpha)/I_{\sigma^+}(\beta)$ for $\lambda_{exc} \sim 532\ nm$ is more than $10^3$, for $\lambda_{exc} \sim 633\ nm$ the ratio $\sim 10$ (both calculated at the peak emission wavelength). Thus the PL contrast is also significantly reduced as $\lambda_{exc}$ approaches the emission maxima. For $\lambda_{exc} \sim 532\ nm$, energy states away from the band-edges of both the conduction and valence band are associated with the electron-hole pairs. De-excitation from the initial states to the band edge and finally populating the excitonic states thus involves strong coupling with the phonons within a valley rather than inter-valley scattering. Secondly, the $LA(K)$ mode is much stronger over the $\alpha$ domains, especially at the edges (figure S10 in the SI) compared to that at the $\beta$ domains. As discussed earlier, the phonons associated with the $LA(K)$ Raman mode are responsible for intervalley scattering and thus decrease valley depolarization, as observed here, i.e. the stronger $LA(K)$ mode at the $\alpha$ domains is correlated with lower local DCP value. Further, deconvolution of the PL spectrum from the $\beta$ domain show relatively higher trion contribution than at the $\alpha$ domains, which have higher exciton contribution. The observation is again commensurate with the relatively higher carrier density in the $\beta$ domains, for flakes on Au. Though local defects and disorder can modify the DCP significantly, other non-radiative excitonic recombination and availability of phonons also modify the $\tau_e/\tau_v$ ratio, hence affect the DCP value. Proximity of the WS$_2$ sample with Au-substrate may also introduce parallel non-radiative relaxation pathways for photo-generated carriers,[52] which may lead to a significant reduction in exciton lifetime hence affect the DCP. Indeed, any potential valleytronic applications with TMDCs will involve an optimal hierarchy of three timescales $\tau_e$, $\tau_v$ and a charge transport (drift-diffusion) timescale $\tau_t$. Note $\tau_t$ is not the same as the carrier scattering timescale that decreases with increasing disorder. Evidently, for simultaneous preservation of valley polarization and realization of high photovoltage upon illumination, the associated timescales should be in the order $\tau_t \ll \tau_e \ll \tau_v$. The coincidence of high DCP regions with regions of lower disorder ($\beta$ domains) is not only consistent with the dynamics of the system i.e.

phonon and impurity scattering, but also lays out the design principles in realizing optically stimulated valleytronics applications.

CONCLUSION

In conclusion, this investigation explored coexisting, spatial variation of optical, spectroscopic, topographic, electrical and electronic properties of 3-fold symmetric domain segmented $WS_2$ flakes, grown on Si by chemical vapour deposition. It maps out coupled physical interactions that nucleate segmentation within single flakes to evolve a correlated picture, elucidating the common origins of the different physical properties and their underlying processes, especially highlighting the role of defects and substrates. Along with evidencing the role of defects in spatial variation of photoluminescence intensity and its energetics, we show that the domain contrast is quenched upon annealing $WS_2$ flakes in sulphur rich environment. Importantly, the vanishing domain contrast is evidenced not only in photoluminescence but also in local conductance, phase imaging and even spatial variation of Raman modes that are sensitive to native defects. Thus, unequivocally tying the origin of domain segmentation to sulphur vacancies in the system. The role of the substrate became apparent once the as-grown flakes were transferred onto Au and reinvestigated. In spite of the uniform topography of as grown $WS_2$ flakes on Si, once transferred to Au the segmented flakes show higher surface roughness along with symmetric height variation. Spatially varying electrical and optical properties in transferred samples directly demonstrate the difference in interactions between high and low electron density regions with the substrate and how that affects the resulting optical and spectroscopic properties of the flakes. Notably, on Au substrate the electron rich $\alpha$ domains exhibit carrier depletion far stronger than the $\beta$ domains, resulting in inverting the spatial correlation between the spectroscopic and electrical properties, across flakes on Au and Si substrates. The substrate induced spatial variation in carrier density is reflected in the spectral composition of photoluminescence, that varies the relative abundance of trion formation across the segments. Evolution of the characteristic Raman modes and their differences across the domains, on either substrates and pre- and post- annealing in Sulphur were analyzed to develop a comprehensive picture of the phonon modes which are crucial to preserving valley polarization in these systems. Finally, photoluminescence measurements quantifying spectrally resolved degree of circular polarization demonstrate that regions harboring higher defect density, thus having higher photoluminescence efficiency, do not preserve valley polarization, in contrast to regions of lower defect density.

EXPERIMENTAL METHODS:

**Growth and transfer of $WS_2$:** The $WS_2$ flakes were controllably synthesized on a conducting Si substrate with a naturally grown surface oxide, and also on $SiO_2$/Si (285 nm $SiO_2$ on Si ) substrate, by chemical vapour deposition.[45] $WO_3$ powder drop casted on cleaned substrate provided the tungsten source, and 500 mg of sulphur powder kept at the upstream end of the quartz tube provided the sulphur. Growth proceeded in argon atmosphere at 900 °C. Some of the as-grown $WS_2$ flakes were subsequently transferred to 50 nm Au coated Si wafer by PMMA based wet transfer technique.

**Photoluminescence and Raman studies:** Spatially resolved PL and Raman studies were conducted at room temperature (RT) with HORIBA XPLORA PLUS micro-Raman setup. The PL and Raman mapping were acquired with 532 nm excitation focused on the sample through a 100X objective with NA ~ 0.9, resulting in spatial resolution of ~ 720 nm. All PL spectra were recorded with 600 grooves/mm grating and 2400 grooves/mm grating was used for conducting the Raman studies. Spectra were recorded using a CCD camera cooled to $-60\ °C$. The typical laser power was used $\sim 1\ mW$ for both PL and Raman experiments.

**AFM studies:** All AFM based experiments were conducted using a Bruker Multimode 8 AFM in the ambient. Topography and phase map were recorded in TM AFM with silicon probe from MikroMasch (HQ: NSC15/Cr-Au BS) having a resonance frequency ~ 325 kHz, stiffness of ~ 40 N/m and tip radius ~ 8 nm. The current maps, conductance ($dI/dV$) maps along were recorded in the conducting AFM (CAFM) mode, equipped with a preamplifier (TUNA2) and gain of $10^{11} V/A$, with Cr/Au coated silicon cantilever having force constant ~ 0.1 N/m and probe diameter ~ 35 nm. During the measurement bias was applied to the sample keeping the cantilever virtually grounded. The $dI/dV$ maps were recorded by modulating the dc sample bias with a small ac-signal ($V_{ac} < 5\% \, V_{dc}$) and detecting in phase ac output voltage using a built-in lock-in amplifier.[66] Spatially resolved photoconductivity studies were conducted with the tip-sample junction flooded with monochromatic radiation of 532 nm from a multimode optical fibre held ~ 5 mm away from the junction.

**Sulfur annealing:** Sulfur annealing of as-grown samples was conducted inside a CVD furnace at a temperature ~ 650°C for 30 min under continuous Argon flow.

*DCP Measurements:* DCP measurements were carried out using a home built HRPL set-up consisting of a Horiba-Jobin Yvon iHR320 spectrometer with a 300 gr/mm grating, a 532 nm diode laser, a 633 nm He-Ne laser in a confocal geometry. The laser was focused on the sample using a 50X objective with NA ~ 0.9, which gave an excitation spot size of ~1 µ$m$ diameter. Laser power was kept constant throughout the experiment. The experiments were performed at RT

ASSOCIATED CONTENT:

**Supporting Information**

Table of the various WS$_2$ Samples; AFM images of homogeneous WS$_2$ flake; Data of AFM, PL, Raman studies on various phase segmented WS$_2$ on Si and Au; DCP at 532 nm excitation. (PDF)

AUTHOR INFORMATION

**Corresponding authors**

*Email: arijit17@iisertvm.ac.in

*Email: j.mitra@iisertvm.ac.in

**Authors' contributions**

A. K., P. K. B., P. V. S. performed the experiment. All authors contributed to writing the manuscript.

**Notes**

The authors declare no competing financial interest.

ACKNOWLEDGEMENT

Authors acknowledge Mr. Renjith N for assistance in sample preparation. JM acknowledges financial support from SERB, Govt. of India (CRG/2019/ 004965), UGC-UKIERI 184-16/2017 (IC) and the Royal Academy of Engineering, Newton Bhabha Fund, UK (IAPPI_77). RNK acknowledges the funding support from the Science and Engineering Research Board, Department of Science and Technology, India, through the Research Grant No.s CRG/2019/004865. AK acknowledges the Ph.D.

fellowship from IISER Thiruvananthapuram. MMS acknowledges financial support from Department of Science and Technology (DST/TMD/HFC/2k18/136), Govt. of India.REFERENCES

1. Zhao, W.; Ghorannevis, Z.; Chu, L.; Toh, M.; Kloc, C.; Tan, P.-H.; Eda, G., Evolution of Electronic Structure in Atomically Thin Sheets of $WS_2$ and $WSe_2$. *ACS Nano* **2013,** *7* (1), 791-797.

2. Mak, K. F.; Lee, C.; Hone, J.; Shan, J.; Heinz, T. F., Atomically Thin $MoS_2$: A New Direct-Gap Semiconductor. *Physical Review Letters* **2010,** *105* (13), 136805.

3. Molas, M. R.; Nogajewski, K.; Slobodeniuk, A. O.; Binder, J.; Bartos, M.; Potemski, M., The optical response of monolayer, few-layer and bulk tungsten disulfide. *Nanoscale* **2017,** *9* (35), 13128-13141.

4. Zhu, B.; Zeng, H.; Dai, J.; Gong, Z.; Cui, X., Anomalously robust valley polarization and valley coherence in bilayer $WS_2$. *Proceedings of the National Academy of Sciences* **2014,** *111* (32), 11606.

5. Barman, P. K.; Sarma, P. V.; Shaijumon, M. M.; Kini, R. N., High degree of circular polarization in $WS_2$ spiral nanostructures induced by broken symmetry. *Scientific Reports* **2019,** *9* (1), 2784.

6. Fan, X.; Jiang, Y.; Zhuang, X.; Liu, H.; Xu, T.; Zheng, W.; Fan, P.; Li, H.; Wu, X.; Zhu, X.; Zhang, Q.; Zhou, H.; Hu, W.; Wang, X.; Sun, L.; Duan, X.; Pan, A., Broken Symmetry Induced Strong Nonlinear Optical Effects in Spiral $WS_2$ Nanosheets. *ACS Nano* **2017,** *11* (5), 4892-4898.28

61. Carvalho, B. R.; Wang, Y.; Mignuzzi, S.; Roy, D.; Terrones, M.; Fantini, C.; Crespi, V. H.; Malard, L. M.; Pimenta, M. A., Intervalley scattering by acoustic phonons in two-dimensional $MoS_2$ revealed by double-resonance Raman spectroscopy. *Nature Communications* **2017,** *8* (1), 14670.

62. Jeong, T.-Y.; Bae, S.; Lee, S.-Y.; Jung, S.; Kim, Y.-H.; Yee, K.-J., Valley depolarization in monolayer transition-metal dichalcogenides with zone-corner acoustic phonons. *Nanoscale* **2020,** *12* (44), 22487-22494.

63. Carmiggelt, J. J.; Borst, M.; van der Sar, T., Exciton-to-trion conversion as a control mechanism for valley polarization in room-temperature monolayer $WS_2$. *Scientific Reports* **2020,** *10* (1), 17389.

64. Shinokita, K.; Wang, X.; Miyauchi, Y.; Watanabe, K.; Taniguchi, T.; Matsuda, K., Continuous Control and Enhancement of Excitonic Valley Polarization in Monolayer $WSe_2$ by Electrostatic Doping. *Advanced Functional Materials* **2019,** *29* (26), 1900260.

65. Miyauchi, Y.; Konabe, S.; Wang, F.; Zhang, W.; Hwang, A.; Hasegawa, Y.; Zhou, L.; Mouri, S.; Toh, M.; Eda, G.; Matsuda, K., Evidence for line width and carrier screening effects on excitonic valley relaxation in 2D semiconductors. *Nat Commun* **2018,** *9* (1), 2598.

66. Bandopadhyay, K.; Mitra, J., Spatially resolved photoresponse on individual ZnO nanorods: correlating morphology, defects and conductivity. *Scientific Reports* **2016,** *6* (1).
37

# Supporting Information

1. Details of the various WS$_2$ Samples discussed in the manuscript.

| # | Sample | Substrate | Remark | Figure # |
|---|--------|-----------|--------|----------|
| 1 | S1 | Si | As-grown | Figure 1 |
| 2 | S2 | Si | As-grown | Figure S2 |
| 3 | S3 | Si | As-grown | Figure S2 |
| 4 | S4 | Si | As-grown | Figure 5, Figure S2, Figure S5, Figure S6 |
| 5 | S5 | Au | Transferred | Figure 2, Figure 3, Figure 5, Figure 7, Figure S3, Figure S5, Figure S9, Figure S10 |
| 6 | S6 | Au | Transferred | Figure S4d, Figure S9 |
| 7 | S7 | Au | Transferred | Figure S4a, Figure S8 |
| 8 | S1A | Si | Annealed | Figure S6 |
| 9 | S2A | Si | Annealed | Figure 6, Figure S7 |
| 10 | S3A | Si | Annealed | Figure S6 |
| 11 | S4A | Si | Annealed | Figure S6 |

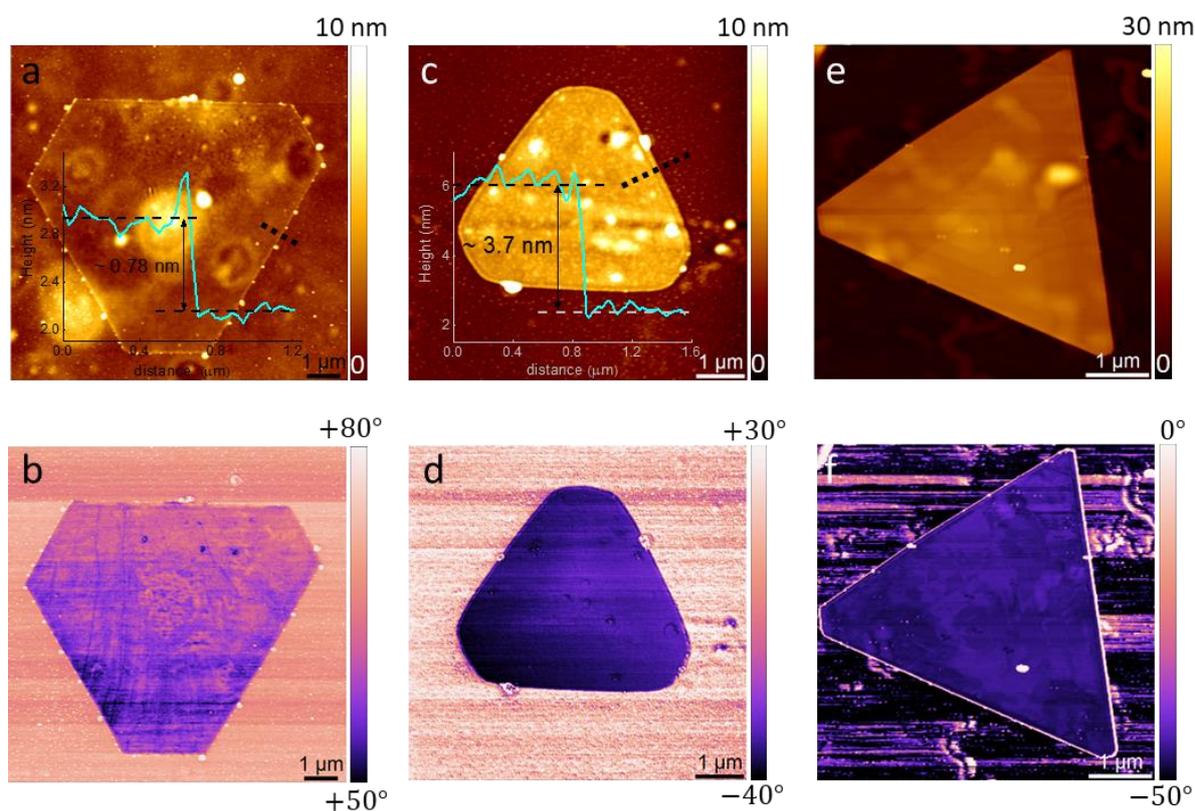

**Figure S1.** AFM Topography and phase maps of CVD grown WS$_2$ flakes of various morphologies (a,b) monolayer (c,d) few monolayer thick and (e,f) multilayer spiral. Majority of the CVD grown samples investigated here appear as above, that is they do not show any signature of symmetric domain segregation.

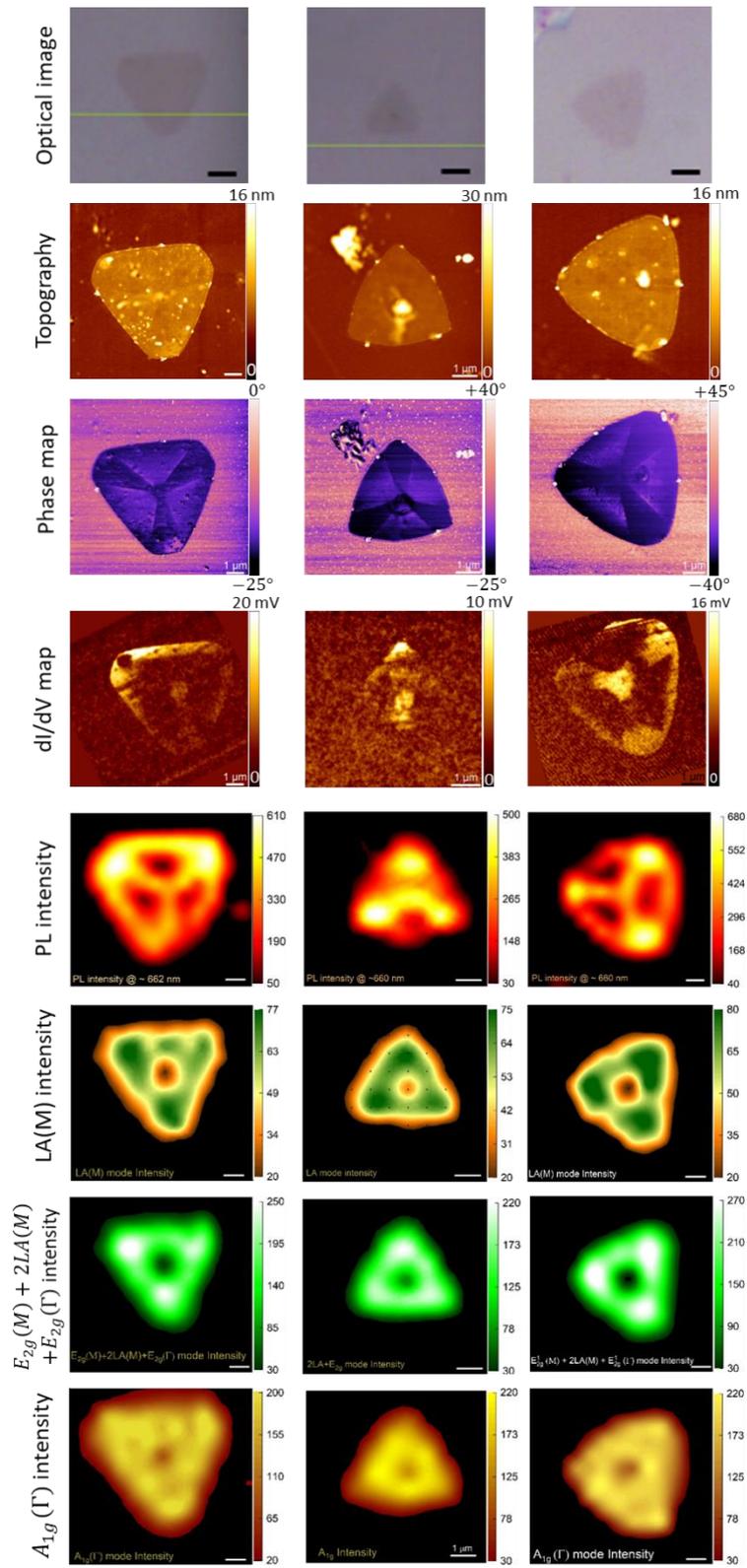

**Figure S2.** Correlated spatially resolved investigations on different WS$_2$ flakes on Si (Samples S2, S3, S4) all showing domain segregation.

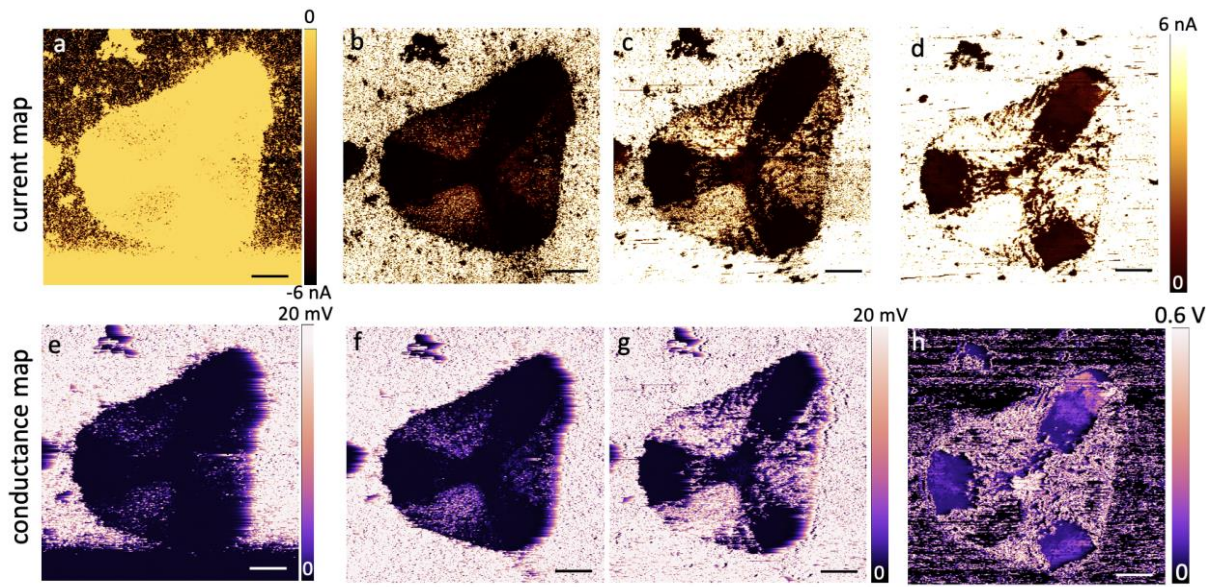

**Figure S3.** (a-d) Current map of WS$_2$ flake (sample S5) recorded at $-1.0\,V$, $+1.0\,V$, $+1.4\,V$, $+2.0\,V$ applied sample bias respectively. (e-h) are corresponding $dI/dV$ maps recorded simultaneously with current map. Scale bar: $2\,\mu m$

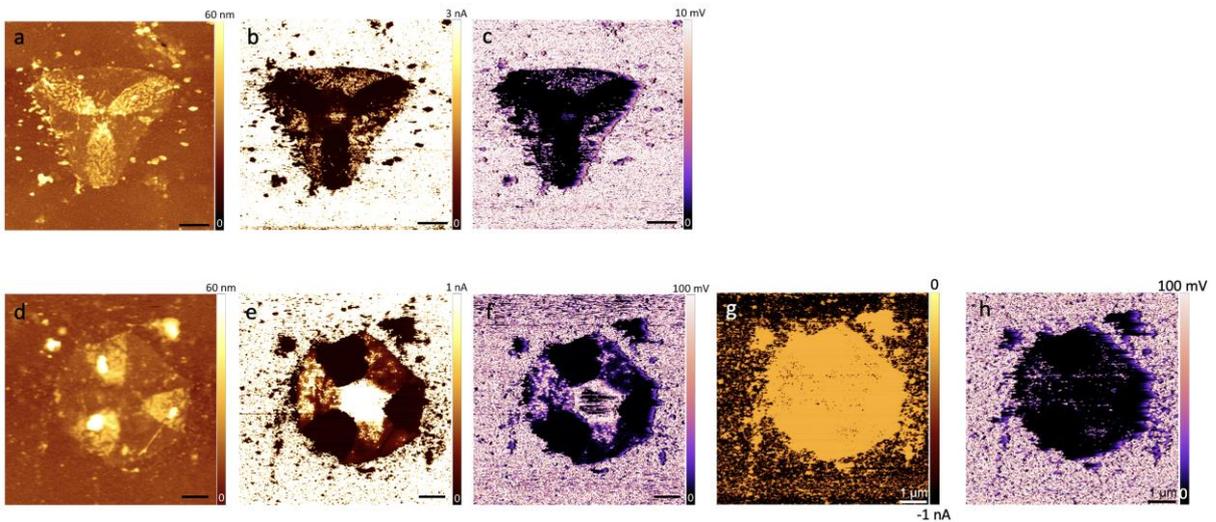

**Figure S4.** (a) Topography, (b) Current map, (c) $dI/dV$ map of a WS$_2$ flake (sample S7) simultaneously recorded at $+1.4\,V$ sample bias respectively. Scale bar: $2\,\mu m$. (d) Topography, Current map at (e) $+1.5\,V$, (g) $-1.5\,V$ ; dI/dV map at (f)$+1.5\,V$, (h) $-1.5\,V$ of an another WS$_2$ flake (sample S6) recorded simultaneously. Scale bar: $1\,\mu m$.

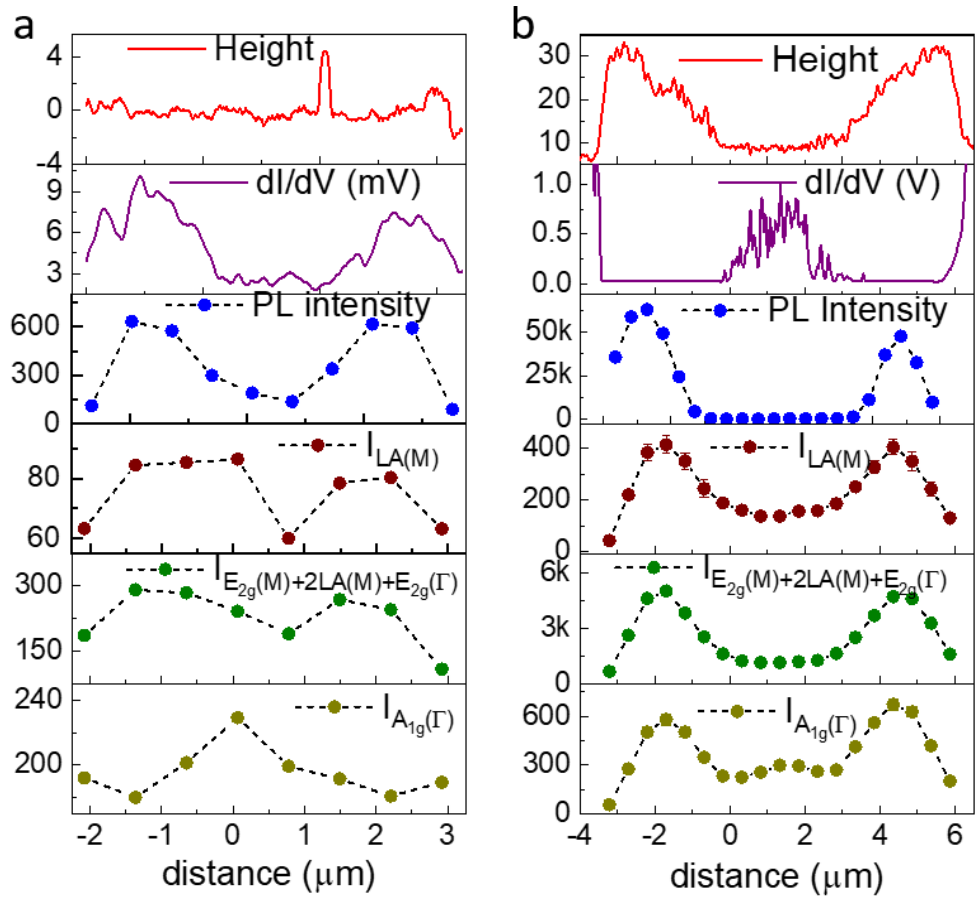

**Figure S5.** Correlated line plots showing variation of intensity of different Raman modes, along with variation in height, conductance and PL intensity for (a) WS$_2$ on Si (sample S4), (b) WS$_2$ on Au (sample S5), along the white-dotted line shown in figure b extends from one $\alpha$-domain to another $\alpha$-domain through a $\beta$- domain.

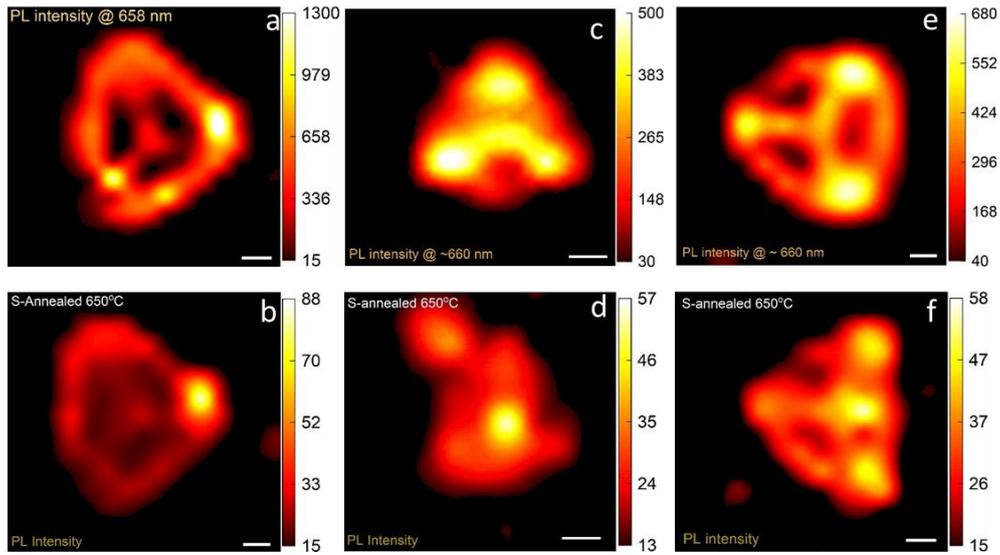

**Figure S6.** PL intensity maps showing domain segregation on three $WS_2$ flakes (a-b) sample S1, (c-d) sample S3, (e-f) sample S4. The first and second row represents the pre- and post- annealing effect on PL emission in Sulphur environment at 650°C, respectively. Scale bar: 1 $\mu m$

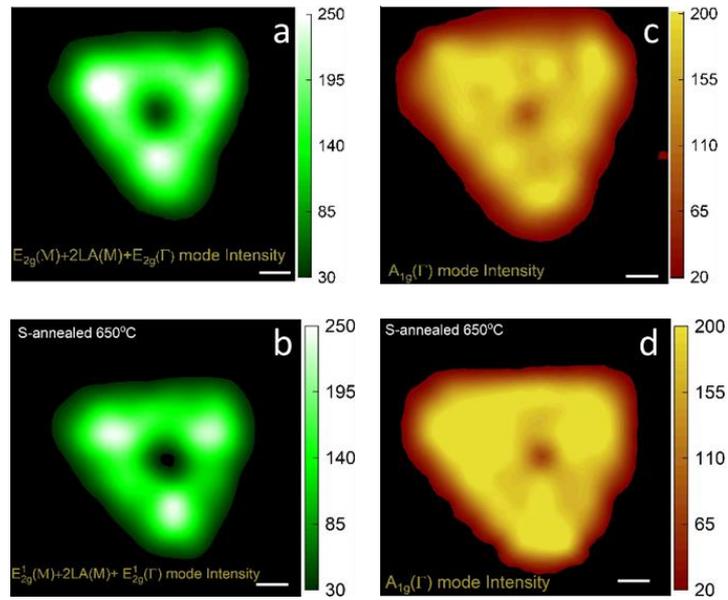

**Figure S7.** Intensity map of (a-b) $E_{2g}^1(M) + 2LA(M) + E_{2g}^1(\Gamma)$, (c-d) $A_{1g}(\Gamma)$ Raman modes for sample S2, before and after S-annealing.

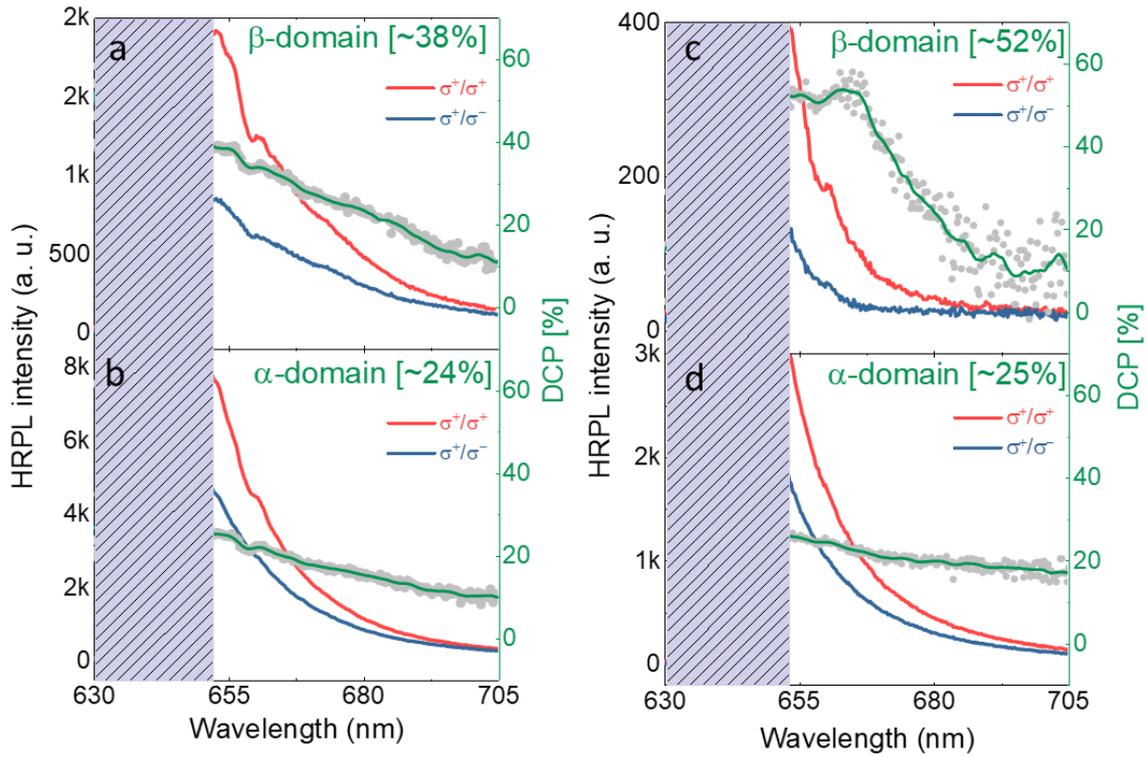

**Figure S8.** PL and DCP from the $\alpha$ and $\beta$ regions for two WS$_2$ on Au, S6 (a-b), S7 (c-d) under RCP ($\sigma^+$) excitation at 633 nm.

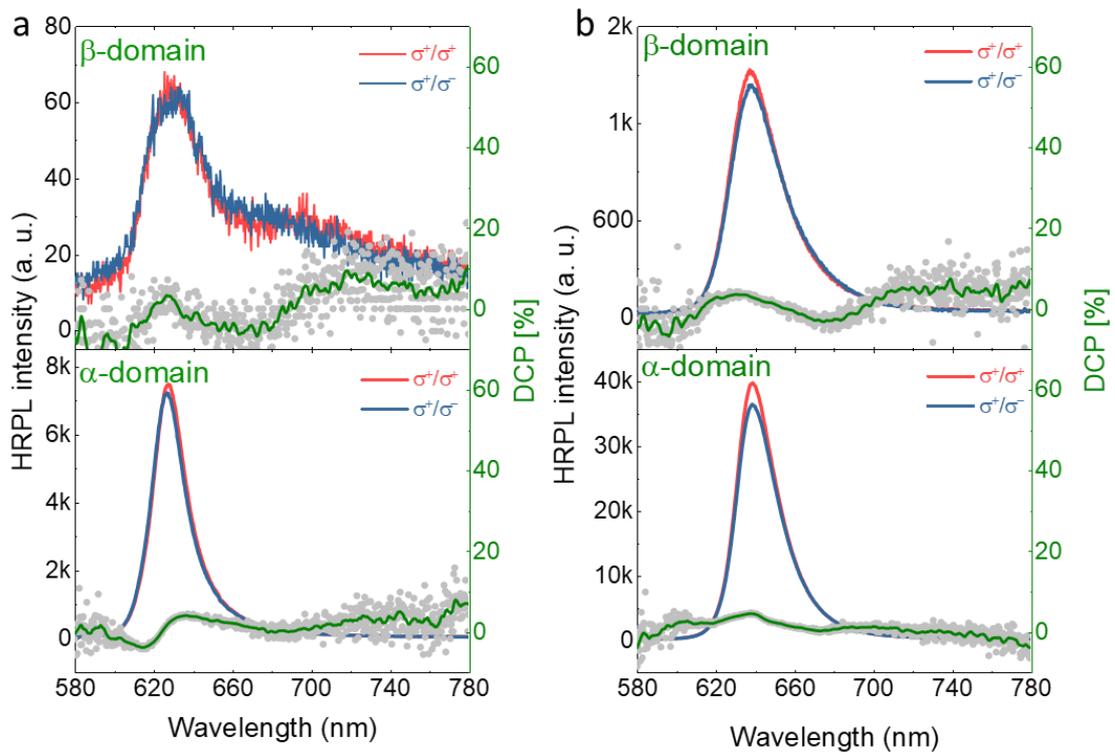

**Figure S9.** PL and DCP spectra from WS$_2$ on Au samples S5 and S6 for RCP ($\sigma^+$) 532 nm excitation.

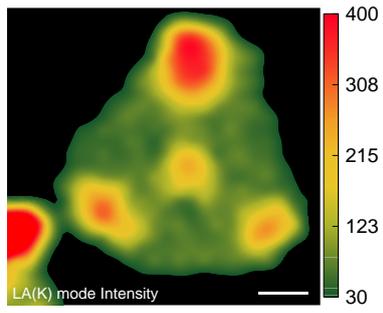

**Figure S10.** Spatial variation of the $LA(K)$ Raman mode intensity across the WS$_2$ flake on Au under 532 nm excitation